\documentclass[conference]{IEEEtran}

\usepackage{amsmath}    
\usepackage{graphicx}   
\usepackage{verbatim}   
\usepackage{color}      
\usepackage{subfigure}  
\usepackage{algorithmicx}

\ifCLASSINFOpdf
\else
\fi

\begin{document}

\title{A Framework for BGP Abnormal Events Detection}

\author{\IEEEauthorblockN{Anisa Allahdadi}
\IEEEauthorblockA{INESC TEC, Faculty of Engineering\\
              University of Porto, \\
              Porto, Portugal \\
              anisa.allahdadi@inescporto.pt}
\and
\IEEEauthorblockN{Ricardo Morla}
\IEEEauthorblockA{INESC TEC, Faculty of Engineering\\
              University of Porto, \\
              Porto, Portugal \\
              ricardo.morla@fe.up.pt}
\and
\IEEEauthorblockN{Rui Prior}
\IEEEauthorblockA{Instituto de Telecomunicações\\
              University of Porto\\
              Porto, Portugal \\
              rprior@dcc.fc.up.pt}
}

\maketitle

\begin{abstract}

Detection of abnormal BGP events is of great importance to preserve the security and robustness of the Internet inter-domain routing system. In this paper, we propose an anomaly detection framework based on machine learning techniques to identify the anomalous events by training a model for normal BGP-updates and measuring the extent of deviation from the normal model during the abnormal occasions. Our preliminary results show that the features generated and selected are capable of improving the classification results to distinguish between anomalies and normal BGP update messages. Furthermore, the clustering results demonstrate the effectiveness of formed models to detect the similar types of BGP anomalies. In a more general context, an interdisciplinary research is performed between network security and data mining to deal with real-world problems and the achieved results are promising.

\end{abstract}

\IEEEpeerreviewmaketitle

\section{Introduction}

Border Gateway Protocol (BGP) is being considered as the primary inter-domain routing protocol which maintains the connectivity of the distinct segments of the Internet known as autonomous systems (ASs). BGP is the protocol of maintaining the table of core routing decisions for different \textit{prefixes} indicating the reachability between ASs. BGP routers send the updates of AS path modifications to its peer ASs and the changes occur based on path, network policies or rule sets. So BGP is basically not a routing protocol but a reachability protocol and since BGP only sends update information when some changes taking place in the network topology or routing policy, it is considered as an incremental protocol rather than periodical protocol. Consequently, analyzing the intensity of BGP updates makes sense on keeping the whole network more stable and properly functioning. 

BGP has enabled the Internet to efficiently become a decentralized system. Therefore the stability and robustness of the BGP is crucial to achieve the global communication goal by preserving the connectivity of the entire network.
However, various abnormal events such as earthquakes, power outage, misconfiguration, extensive worm attacks or other malicious network activities, disable the reachability and stability of particular parts of the network regionally or globally. Such abnormal events affect the routing infrastructure and cause delays, data loss and connectivity problems over the Internet. Consequently, early detection of such sort of anomalies and misbehaviors is of prominent significance to ensure the stability, availability, and efficiency of the Internet.

In this study we present an anomaly detection framework based on machine learning algorithms applied to well-known BGP abnormal events. The anomaly detection systems, generally, try to make a model of normal behavior and announce any deviation from that normal model as an anomaly. However, in this process there exist false alarms correspond to harmless activities or less serious failure in the system being signaled as anomalies. Therefore, the main objective of every anomaly detector is to discover all possible abnormal activities while minimizing the ratio of the false alarms. The proposed anomaly detection framework in this study consists of three main steps: feature extraction, feature selection, and feature generation. First a set of features from BGP messages are produced, then the most relevant features are selected, and eventually new features are generated based on the correlation of the selected features. The feature generation part deals with the learning and classification process and utilizes the machine learning algorithms, namely Support Vector Machines (SVM).

The rest of the paper is organized as follows. Some background studies and related works are briefly introduced in section two. Section three explains the system architecture and illustrates the main approaches applied in the framework. In section four experimental setup and results are presented. Finally section five concludes the paper and addresses some future lines of work.

\section{Related Works}

Several studies have been conducted on network data to discover BGP abnormal events. In an early line of work Li et al \cite{Ref2} proposed an Internet Routing Forensics framework to process BGP routing data and extract rules of abnormal BGP events. They applied data mining techniques and train the framework to learn the rules for different types of BGP anomalies and showed that these rules are effective in detecting the occurrences of the similar events. They utilized a feature selection method to pick the features with the highest information gain. The selected features in some cases match the selected features of our method. However they select 9 features among 35 for all the BGP events, while the set of our selected features vary from dataset to dataset in addition to a number of new features presented based on pairwise correlation of the features.

In a similar approach to \cite{Ref2}, Cazenave et al \cite{Ref1} applied data mining algorithms to learn from labeled abnormal data and distinguish unseen BGP events. They have extracted several numerical features for certain time bins and leveraged a number of classifications algorithms, among which SVM had outperformed the others, to detect similar type of events in famous BGP events datasets. In another line of work, Dou et al. \cite{Ref3} automatically formed the hierarchy of abnormal BGP events by devising a clustering method and obtained a set of classification rules which enabled them to label unknown BGP events to the most similar category. However, the proposed methodology is not always able to distinguish the exact category in the lowest level of the hierarchy or even differentiate the normal data from abnormal events. In this study we obtained more precise results in terms of detection of abnormal BGP events compared to the presented results in \cite{Ref3}.
 
Furthermore, Zhang et al \cite{Ref4} proposed an instant-learning framework recognizing anomalies based on their deviation from the normal dynamics of BGP updates. They applied wavelet transforms to reveal temporal structure of update messages and utilize clustering algorithms to distinguish between normal profiles and outliers.

\section{Data Descripton}
This section describes the origin of the datasets, their structure and other relevant aspects of the data.

\subsubsection{Datasets}
In this work we focus on six well-known datasets containing abnormal BGP events. Three of them (Nimda, Slammer and CodeRed) correspond to Internet worm outbursts, whereas the other three (Eastcoast, Florida and Katrina) represent extensive blackouts. Each dataset consists of 60 days of observation, starting approximately 30 days before the beginning of the anomaly. The duration of anomalies themselves varies from 16 hours to 4 days, as shown in \ref{table1}. Since there is no publicly available data from a single remote route collector (RRC) spanning all six events, Table \ref{table1} also shows the RRC where dataset originated.
 
    \begin{table} [h]
      \caption{List of BGP abnormal events} 
      \centering 
      \begin{tabular} [|c|c|c|c|]{|p{2.1cm}| p{1.8cm}| p{1.5cm}| p{1.8cm}|} 
  	\hline
\textbf{Event Name} & \textbf{RRC} & \textbf{Date}	& \textbf{Duration}\\
  	\hline
Nimda Worm & rrc04, Geneva & 18-Sep-2001 & 2 days\\
  	\hline
Slammer Worm & Routviews, Oregon & 25-Jan-2003 & 1-2 days\\
  	\hline
Codered Worm & rrc02, Paris & 19-Jul-2001 & 16 hours\\
\hline
East-coast  Blackout &	Routviews, Oregon &	14-Aug-2003	& 2 days\\
  	\hline
Florida Blackout & rrc11, USA & 03-Sep-2004 & 4 days (1st day)\\
  	\hline
Katrina Blackout & rrc14, USA & 29-Aug-2005 & 4 days (1st-2nd days)\\
  	\hline
  \end{tabular}
  \label{table1}
  \end{table}

\subsubsection{Features List}

The raw BGP message data is retrieved from a subset of the update archives of RoutViews \cite{Ref6} and RIPE \cite{Ref7}.
The events from the raw datasets are grouped into one minute bins, and data conversion process, field filtering, duplicate removal, and event measurement are performed on the raw events. These steps are described in detail in \cite{Ref8}.
From the dataset, we extracted a number of features that are useful for the identification of routing events and anomaly detection.
For instance, as the routers explore alternative paths due to a routing event such as link failure or topological modification, the number of BGP updates messages may change drastically. Such alterations contribute to significant data features to build various perspectives of routing dynamics that facilitate the use of machine learning algorithms for anomaly detection.

The learning algorithm is applied to a set of 18 primary features, listed in Table \ref{table2}. The number of BGP announcements and withdrawals (parameters 1--3) are indicative of the Internet routing dynamics.
Since one BGP update message can involve multiple prefixes and vice versa, there can be multiple updates for a given prefix in a single bin. The number of updated prefixes is considered in parameters 4--6. As Labovitz et al.\ discussed in \cite{Ref9}, routing instabilities might be classified as forwarding instability, policy fluctuation, and pathologic (or redundant) updates. In the sequence of BGP update messages some types of successive events are defined which represent instabilities belong to those categories (parameters 7--15).  The above mentioned parameters are described in more detail in \cite{Ref2}.

The number of reachable prefixes can be affected by large scale anomalous events. Therefore, an additional feature we used (parameter 18) that is the number of reachable prefixes in each time bin. However, some prefixes can be reachable through a single neighboring AS, meaning that they become unreachable if, for some reason, the connection to the BGP router(s) in that neighboring AS is lost. The number of active eBGP peers was used as an additional feature (parameter 17) to make it easier for the learning algorithm to deal with this effect. Finally when a new eBGP peer is added or a connection to a previously lost eBGP peer is restored, the routing tables must be exchanged implying a large number of updates. Thus we added a feature containing the number of peers performing table transfers in the bin period, estimated from the logs (state changes and updates).

\begin{figure*}
\centering
\includegraphics[scale=.62]{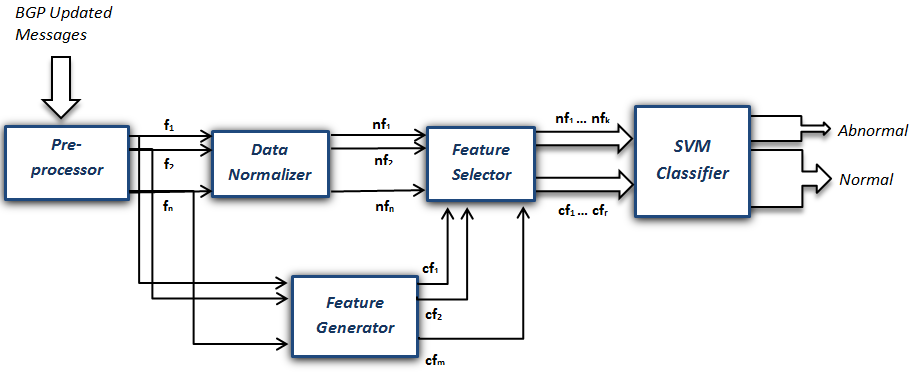} 
\caption{The architecture of the proposed framework}
\label{fig1}
\end{figure*}

    \begin{table}
      \caption{BGP related features' list} 
      \centering 
      \begin{tabular} [c]{ |p{0.5cm}| p{1.8cm}| p{5.0cm}|} 
  	\hline
\textbf{ID} & \textbf{Features} &	\textbf{Description}\\
  	\hline

1 & Announce & \# of announcements\\
2 & Withdrawal & \# of withdrawals\\
3 & Update & \# of updates (=Announce+Withdrawal)\\
4 & AnnouPrefix & \# of announced prefix\\
5 & WithdwPrefix & \# of withdrawn prefix\\
6 & UpdatedPrefix & \# of updates prefixes (=AnnouPrefix+WithdwPrefx)\\
7 & WWDup & \# of duplicate withdrawals\\
8 & AADupType1 & \# of duplicate announcements (all fields are the same)\\
9 & AADupType2 & \# of duplicate announcements (only AS-PATH and NEXT-HOP are the same)\\
10 & AADiff & \# of new-path announcement (implicit withdrawals)\\
11 & WADupType1 & \# of re-announcements after withdrawing the same path (all fields are the same)\\
12 & WADupType2 & \# of re-announcements after withdrawing the same path (only AS-PATH and NEXT-HOP are the same)\\
13 & WADup & WADupType1 + WADupType2\\
14 & WADiff & \# of new paths announced after withdrawing na old path\\
15 & AW & \# of withdrawals after announcing the same path\\
16 & NPeers & \# of peers\\
17 & ReachPrefix & \# of BGP Atoms\\
18 & TblExchgA & table exchange status\\
  	\hline	
  \end{tabular}
  \label{table2}
  \end{table}

\section{System Architecture}

This section describes the architecture of the proposed framework, its components and their operations. The main functionality of the system is to select and generate the best set of features with the highest correlation with the anomalous data in its occurrence period. Figure \ref{fig1} demonstrates the various modules of the framework and the interaction between them. The input data are basically BGP update messages on which several pre-processing operation have been already performed in \cite{Ref8}. The emerging features of the pre-processor component are the main features for this work and the rest of the data processing and classification algorithms will be described in the following sections.

\subsection{Data Normalizer}

One of the early steps of pre-processing in this work concerns the normalization of the data which is crucial when dealing with parameters of different units and scales. Therefore we applied the standardization on the data using Equation \ref{formula 1} to every features. Having subtracted the mean for each data point and dividing the result by the standard deviation we end up having zero means and unit variance. This assumption is regarded to scale the attribute data in order to fit in a specific range and make the further computations meaningful.

\begin{equation}
x_{normal} = \frac{x- \mu}{ \sigma}
\label{formula 1}
\end{equation}

\begin{figure*}[t]
\centering
\includegraphics[width=0.9\textwidth]{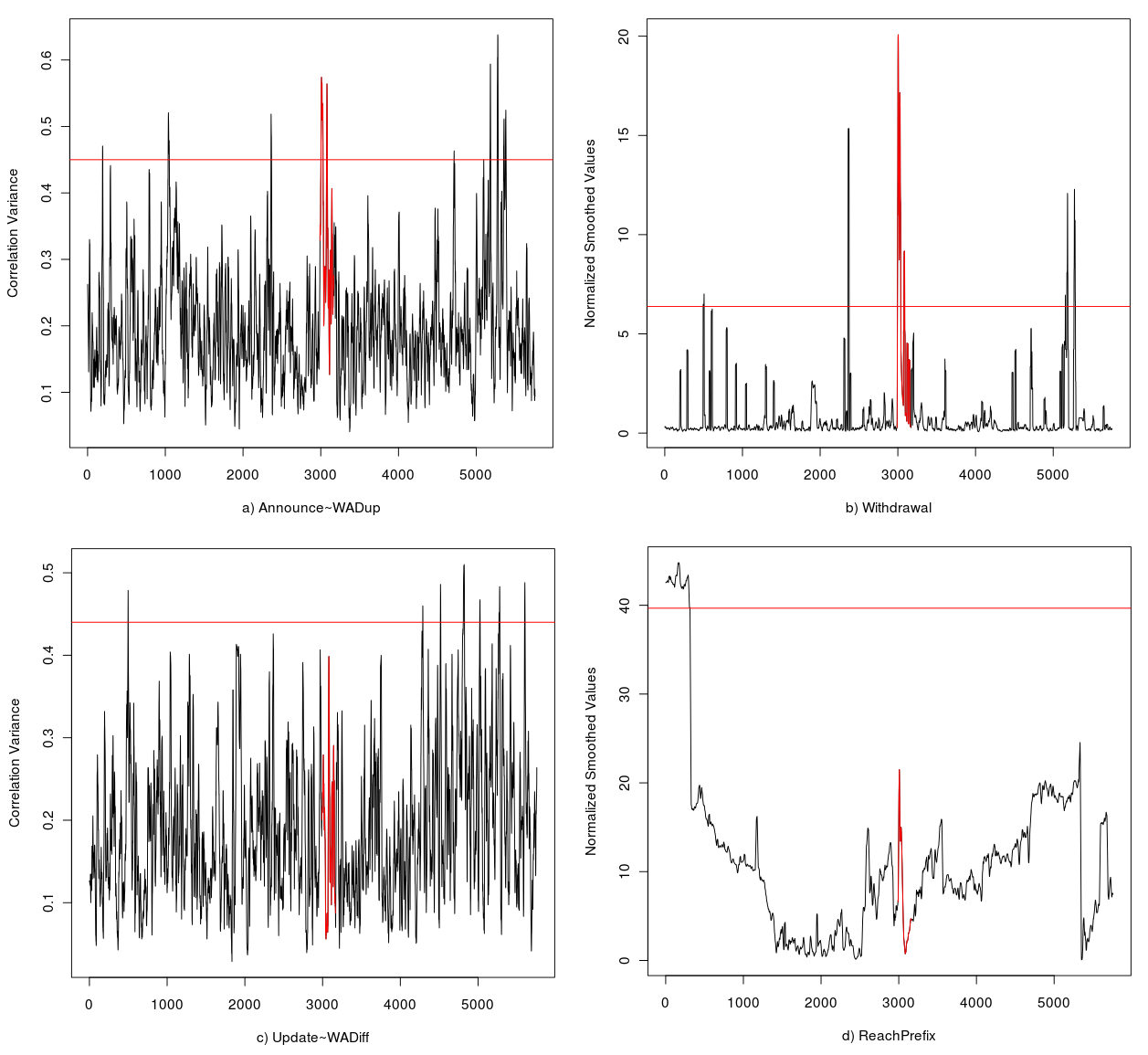} 
\caption{The values of some sample features of Slammer dataset in a 15 minutes time bin. Period of anomaly is marked with red and the horizontal red lines show the abnormality threshold above which abnormal data is detected a) Correlation between Announce and WADup b) Withdrawal c) Correlation between Update and WADiff d) ReachPrefix}
\label{fig2}
\end{figure*}

\subsection{Feature Generator}

This module generates new features based on the correlations between the original features derived from pre-processing step. The motivation behind this work is to have a more accurate anomaly estimation observing how the correlation of variable pairs change during the normal period and when the anomaly occurs.
Correlation is a statistics that measures the degree of relationship between two variables. The value of correlation coefficient, a number between -1 and 1, shows the extent to which two variables are proportional. As the result of correlation does not depend on the measurement unit scales, we feed the original features (before normalization) to this module and obtain the new variables as the pairwise correlations. In this study the Pearson's correlation coefficient is applied and the correlation significance test is performed to calculate the reliability of the correlations (\ref{formula 2}).
Given two event-series x and y, the Pearson's correlation coefficient is defined as:
 
\begin{equation}
r=\dfrac{\sum_{i=1}^{N} {(x_{i}-\mu_{x})(y_{i}-\mu_{y})}}{(N-1) \sigma_{x} \sigma_{y}}
\label{formula 2}
\end{equation}

The significance level calculated for each correlation indicates the quality of the magnitude and depends on the size of the sample from which it was computed. In fact, the significance result of two variables shows the probability that the observed correlation has occurred by chance. Therefore, in larger sample size we expect to observe higher reliability without restricting the significance value to greater numbers. More detail information can be found in \cite{Ref9}. However, in this component, the correlations computed by Pearson's method, has passed the significance test afterwards to make sure the reliability of the correlations and remove the insignificant ones from the output.

\subsection{Feature Selector}

The main activity of this component is to pick the most salient features that show the highest relation with the occurred abnormal BGP event. As Figure \ref{fig1} shows, two sets of variables are provided for this component, first normalized features derived from the pre-processing module and second correlated features as the output of the feature generator module. Any of these variables that reveal the greatest change during the anomaly period and remain unchanged in the normal period, obtain higher score to be selected for the classification phase. Hence, two important aspects is considered to present a feature as the selected one, larger number of detected changes in the time of anomaly, and less false alarms during the normal period. 

To detect the amount of deviation from the normal behavior, it is assumed that the variable has normal distribution and any value beyond the ($ \mu \pm 2\sigma $) is considered as an outlier or abnormality. To calculate the total anomaly score of a variable we perform majority vote to smooth the signal of approved changes, means from every k time bins we select the greater number of Normal or Abnormal occurrences and announce the entire bin as \textit{Normal} or \textit{Abnormal}. Eventually, the candidate features are the ones with more number of abnormalities in the anomaly period and less abnormal samples in the normal periods (confusion matrix). In Figure \ref{fig2} two sets of acceptable and non-acceptable variables are displayed and the range of approved changes for each variable is demonstrated by the red horizontal lines.

\subsection{SVM Classifier}

The SVM classifier is applied on the datasets of well-known BGP events and the aforementioned features, to distinguish between normal data and abnormal events. The model is built based on one-classifier SVM for the purpose of novelty detection. In fact the model is trained with the normal data and is expected to mark abnormal events when there is a mismatch with the normal model. In the next section more results are demonstrated in details reflecting how different modules of the proposed framework operate.

\section{Experimental Results}

In the first set of experiments conducted on six BGP abnormal events, one SVM model is built for each data set separately. The normal models belong to the clean period of each data set before the occurrence of the abnormal event. Then the test set, containing 50\% of unseen normal data and 50\% of abnormal data are passed to the model to verify the ability of detecting normal and abnormal data. Applying one-classification SVM with the parameters identified by the tune method in R for the Slammer dataset provided us with the results presented in Table \ref{table3}.

    \begin{table} [h]
      \caption{Anomaly detection results for Slammer dataset} 
      \centering 
      \begin{tabular} [|c|c|c|c|]{|p{2.8cm}| p{1.4cm}| p{1.4cm}| p{1.2cm}|} 
  	\hline
\textbf{Data Features (n=number of features)} &\textbf{Anomalies Detected (True Positive)} & \textbf{False Alarms} & \textbf{Accuracy}\\
  	\hline
All Features (n=18) & 100\% & 98\% & 51\% \\
  	\hline
Selected Features (n=6) & 82\% & 12\% & 85\%\\
  	\hline
Selected Features + New Features (n=4+4) & 99\% & 3\% & 98\%\\
\hline
  \end{tabular}
  \label{table3}
  \end{table}
  
As previously discussed, the highest number of anomalies detected along with the lowest rate of false alarms is the most promising case which is obtained in the third case represented in Table \ref{table3}. In the first row of the above table, all the normalized features have been selected and the model is trained blindly using the attributes of all these variables. In the second row the model is built based on the selected features and in the third row a combination of selected features and new features are provided to yield the highest precision and lowest false alarms. Combining the selected set of features and the new features as the most prominent features and feeding them to the SVM classifier achieve the recognition rate of 99\% of the anomalies while producing only 3\% of false alarms. Noticeable improvement of the detection rate and false alarms results demonstrate the superiority of the proposed methods in feature selection and feature generation phases.

\begin{table*} 
\caption{Detection rate of BGP events with distinctive datasets} 
\centering 
\begin{tabular} [|c|c|c|c|c|c|c|]{|p{1.9cm}| p{1.6cm}| p{1.6cm}| p{1.6cm}| p{1.6cm}| p{1.6cm}| p{1.6cm}|} 
\hline
\textbf{ } & {Nimda Ds.} & {Slammer Ds.} & {Codered Ds.} & {Eastcoast Ds.} & {Florida Ds.} & {Katrina Ds.}\\
\hline
Nimda Mdl. & 99\% & 100\% & 82\% & 85\% & 70\% & 62\%\\
\hline
Slammer Mdl. & 90\% & 100\% & 81\% & 50\% & 32\% & 50\%\\
\hline
Codered Mdl. & 70\% &  51\%  & 91\% & 45\% & 55\% &  49\%\\
\hline
Eastcoast Mdl. & 50\% & 50\% &  50\% &  100\% & 50\% &  54\%\\
\hline
Florida Mdl. & 51\% & 50\% & 50\% & 41\% & 91\% & 50\%\\
\hline
Katrina Mdl. & 53\% & 77\% & 63\% & 62\% & 60\% &  90\%\\
\hline
\end{tabular}
\label{table5}
\end{table*}

    \begin{table*} 
      \caption{Distance matrix of BGP events} 
      \centering 
      \begin{tabular} [|c|c|c|c|c|c|c|]{|p{1.9cm}| p{1.6cm}| p{1.6cm}| p{1.6cm}| p{1.6cm}| p{1.6cm}| p{1.6cm}|} 
  	\hline
\textbf{ } & {Nimda Ds.} & {Slammer Ds.} & {Codered Ds.} & {Eastcoast Ds.} & {Florida Ds.} & {Katrina Ds.}\\
  	\hline
Nimda Mdl. & 0 & 9 & 38 & 64 & 69 & 74\\
  	\hline
Slammer Mdl. & 9 & 0 & 59 & 100 & 105 & 63\\
  	\hline
Codered Mdl. & 38 &  59  &  0 &  96 &  77 &  69\\
\hline
Eastcoast Mdl. & 64 & 100 &  96 &   0 & 100 &  74\\
\hline
Florida Mdl. & 69 & 105 & 77 & 100 & 0 & 71\\
\hline
Katrina Mdl. & 74 & 63 & 69 & 74 & 71 &  0\\
\hline
  \end{tabular}
  \label{table6}
  \end{table*}

      \begin{table} [h]
        \caption{Anomaly detection results based on the best set of features in all datasets} 
        \centering 
        \begin{tabular} [|c|c|c|c|]{|p{1.8cm}| p{1.4cm}| p{1.4cm}| p{1.4cm}| } 
    	\hline
  \textbf{Data Set} &\textbf{Anomalies Detected (True Positive)} & \textbf{Accuracy} & \textbf{Precision}\\
    	\hline
  Nimda & 96\% & 95\% & 94\%\\
    	\hline
  Slammer & 99\% & 98\% & 97\%\\
    	\hline
  Codered &88\% & 88\% & 88\%\\
  \hline
  East-coast &100\% &97\% &94\%\\
  \hline
  Florida &100\% &83\% &75\%\\
  \hline
  Katrina &97\% &92\% &88\%\\
  \hline
  
    \end{tabular}
    \label{table4}
    \end{table}
    
In Table \ref{table4} detection results are demonstrated for all the BGP events in this experiment. It must be considered that the results are based on the best set of features for each data set. Except from the \textit{Codered} sample, the true positive rate of anomalies detected are extremely high, in some cases close to 100\% that to best of our knowledge is the best results ever acquired. However the \textit{Codered} result is also acceptable to some extent. Other metrics mentioned in Table \ref{table4}, accuracy and precision, provide more precise prospect of the obtained results, both for the true positive rate and false negative frequency. 

A very important aspect of each anomaly detection model is its capacity to be generalized, means in what extent the built model is able to detect the similar new failures in the future. To achieve this goal, we examined every BGP abnormal event towards the other models and measure the distance between pairs of models to perceive the similarity of the six events described earlier.

The outcome of anomaly detection accuracy for each of the aforementioned abnormal events against different models are presented in Table \ref{table5}. The columns show the model built upon specific data sets and the rows entitle the data sets regarding the BGP events. Hence, it is expected to have the most accurate results on the main diagonal, testing a data set for the same model built on top of the identical set of features. However, there is an exception when testing the \textit{Slammer} data against the \textit{Nimda} model which produced the more accurate result for \textit{Slammer} rather than \textit{Nimda} itself. The accuracy matrix provided is clearly not symmetric. Therefore, we utilized a distance measure defined in \cite{Ref10} to capture the mutual fitness of two data sets. In Equation \ref{formula 3}, $AM$ stands for the accuracy matrix and $d$ shows the distance evaluated between abnormal events of $ i $ and $ j $.

\begin{equation}
 d(i,j) = |AM(i,i) + AM(j,j) - AM(i,j) - AM(j,i)| 
 \label{formula 3}
\end{equation}

In this equation, the terms $ AM(i,i) $ and $ AM(j,j) $ indicate the feasibility of the model given its own fitted test set. The cross terms $ AM(i,j) $ and $ AM(j,i) $ show the likelihood of a detection output generated by the fitted model of another data set. If the accuracy value of $ AM(i,j) $ and $ AM(j,i) $ are proportionally high and closely equal, it indicates a very small distance between abnormal events of $ i $ and $ j $. In other cases, when the model $ i $ is best fitted with the data of $ j $ but the model $ j $ do not provide an accurate detection result for data of $ i $, the evaluated distance might be relatively larger. The conclusive distance matrix is shown in Table \ref{table6}.

\begin{figure*} [!t]
\centering
\includegraphics[scale=.5]{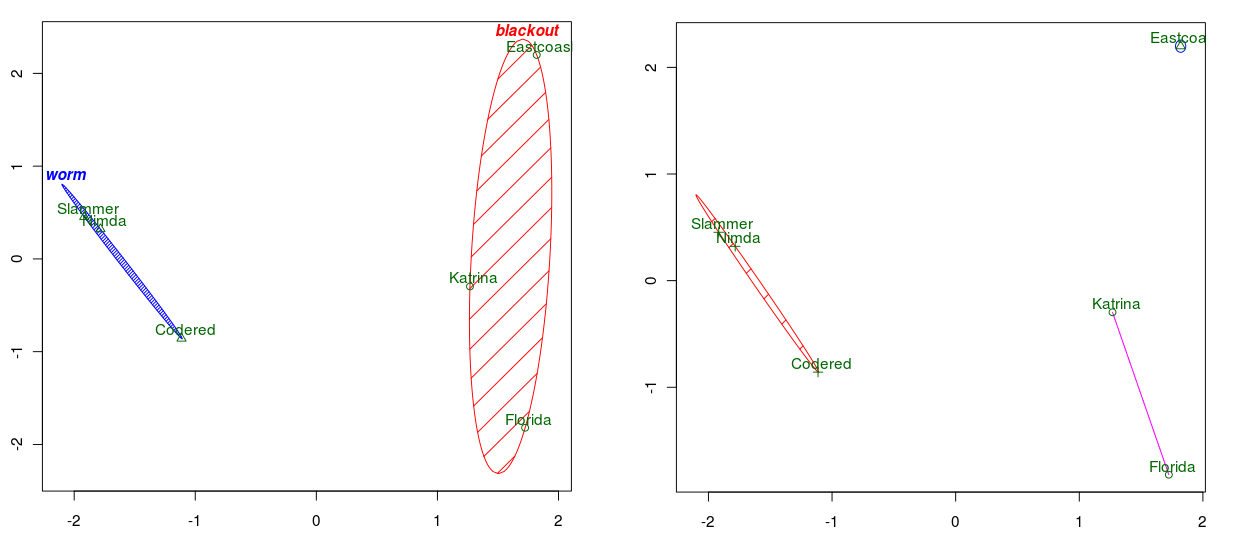} 
\caption{Clustering similar BGP events}
\label{fig3}
\end{figure*}

Having generated the distance matrix of the BGP abnormal events, a clustering method (kmeans in R) is performed on the six data sets to observe how the similar events can be detected by the closest models. Clustering with k=2, precisely provided two separate groups of \textit{worm attacks} and \textit{extensive blackout}. Extending the clusters to k=3, put the \textit{Eastcoast} apart while the rest of the events remained in the same groups. This classification implies that similar new BGP abnormalities are detectable with a bounded error while applying the models of the same type of abnormality without the necessity of having the identical set of features. 

\section{Discussion}
In this work, we analyzed BGP behavior for two types of BGP abnormalities, \textit{worm attacks} and \textit{extensive blackouts}. We extract a number of metrics from the enumeration process performed on BGP update messages and several features are generated from the correlation between variable pairs. The final SVM model is created based on the top ranked set of features for the normal data and are tested for detecting BGP abnormal events for the unseen data. The achieved results regarding the detection accuracy and precision show a considerable improvement over the selected set of features and generated set of features. Output in Table \ref{table4} indicate the prominent results of anomaly detection in BGP events under investigation. 

Due to the fact that BGP anomalies happen rarely in a long period of time and false alarms occur quite often, the results are recommended to be evaluated by network administrating team that might require re-estimation of the results, detected anomalies vs. false positive ratio. However, the results acquired in this study presented the most accurate results ever achieved in this context compared to the state of the art.

An important direction of work is done in the time of detecting one BGP abnormality with a model made from another event data with different set of salient features. The emerging clusters show that BGP events originated from the same cause, worm or blackout, are more likely to be located in the same group and more probable to be detected by similar models. The hierarchy demonstrated in Figure \ref{fig3} has some similarities with the hierarchy of BGP events presented in \cite{Ref3}, for example the close distance of \textit{Nimda} and \textit{Slammer}, these two with the \textit{Codered}, or \textit{Katrina} and \textit{Florida} with each other. However, the clusters of this work makes more sense when putting the same type of BGP events in the same group as it allows the network manager to distinguish the closest model for anomaly detection purposes or the possible root cause diagnoses.

\section{Conclusion and Future Works}

In this work we proposed an anomaly detection framework applied on BGP traffic data based on a feature extraction and generation approaches.
The methodology used for feature generation and selection in this study provided promising results with high precision in anomaly detection and low false positives which outperformed the existing results of BGP anomaly detection both in detection rate and hierarchy estimation of the BGP events.
 
Open issues of this work include improving the obtained results by decreasing the false positive rate while keeping the detection precision high. 


\section*{Acknowledgment}

This work is financed by the ERDF – European Regional Development Fund through the COMPETE Programme (operational programme for competitiveness) and by National Funds through the FCT – Fundação para a Ciência e a Tecnologia (Portuguese Foundation for Science and Technology).

\nocite{*}
\bibliography{biblografia}
\bibliographystyle{IEEEtran}

\end{document}